\documentclass[letterpaper,aps,showpacs,floatfix,11pt,prc]{revtex4-1}
\usepackage{graphicx}
\usepackage{amsmath,amssymb,amsbsy,bm}
\usepackage{graphicx}
\usepackage{comment}
\usepackage{float}
\usepackage[colorlinks=true,linkcolor=blue,citecolor=blue,urlcolor=blue]{hyperref}

\begin{document}

\title{Consistent Implementation of Non-Zero-Range Terms into Hydrodynamics}
\author{Scott Pratt}
\affiliation{Department of Physics and Astronomy and National Superconducting Cyclotron Laboratory\\
Michigan State University, East Lansing, MI 48824~~USA}
\date{\today}


\begin{abstract}
Non-zero-range interactions are often incorporated into mean field theories through gradient terms. Here, a formalism is developed to incorporate such terms into hydrodynamics. These terms alter expressions for the entropy, chemical potential, temperature and the stress-energy tensor. The formalism respects local conservation of energy, charge and entropy. The formalism leads to static solutions where the temperature, chemical potential and hydrodynamic acceleration all vanish, even when the density profile might be non-uniform. Profiles for a phase boundary and for correlation functions are calculated to illustrate the gradient modifications for various thermodynamic quantities. Also, for hydrodynamic calculations that add thermal noise to generate density-density correlations of the desired strength, an additional noise term is derived so that, at equilibrium, correlations are generated with both the correct size and length scale.
\end{abstract}

\maketitle

\section{Introduction}
\label{sec:intro}

Mean field theory, or Landau theory, for thermal systems typically consider the free energy density to be a function of the temperature, density and density gradients \cite{HuangStatMech},
\begin{eqnarray}
f(\bm{r})&=&\bar{f}(T,\rho(\bm{r}))-\frac{\kappa}{2}\rho\nabla^2\rho.
\end{eqnarray}
The gradient term is physically motivated by short, but non-zero-range attractive interactions. For a given particle, the effective number of neighbors within the range of the interaction differs from a uniform background, and if the curvature, $\nabla^2\rho$, is positive the effective number of neighbors increases relative to a uniform density profile and the free energy is reduced. Mean field theory can be applied to numerous systems, such as the liquid-gas transition, where $\rho$ is the density, or to problems in magnetization, where $\rho$ is replaced by the local magnetization. In nuclear physics, mean field pictures have been applied to understanding the surface profiles between phases. This includes the nuclear liquid-gas phase transition \cite{Ravenhall:1984ss}, between the quark-gluon plasma and hadronic phases at zero baryon density \cite{Csernai:1992tj,Csernai:2002bu} (now known not to exist as there is no first-order phase transition), and between chiraly restored and broken phases at high baryon density and temperature \cite{Mintz:2012mz} (hypothesized to exist \cite{Bazavov:2017dus}). In general, such methods can be applied to describe correlations near any critical point \cite{HuangStatMech}, and though the critical exponents are not correct for three-dimensional systems at the critical point, as can be seen by the Ginzburg criteria \cite{HuangStatMech}, should be a good description away from the critical point and should not differ qualitatively even close to the critical point. In nuclear structure calculations gradient terms are also often applied to density functional theory\cite{brink}, where the potential used for the Schr\"odinger equation becomes a function of both the density and the its gradients. Gradient terms have been applied to hydrodynamic treatments of spinodal decomposition for the chiral transition \cite{Steinheimer:2013xxa,Steinheimer:2013gla,Steinheimer:2012gc,Randrup:2010ax} and the liquid-gas transition \cite{Heiselberg:AnnPhys,Heiselberg:1988oha}, but not in a way where all thermodynamic quantities are consistently altered. Similar effects have been taken into account in Boltzmann transport by using a non-zero range to calculate the density used to generate local potentials \cite{Napolitani:2014ima,Chomaz:2003dz,Borderie:2001jg,Colonna:2002ti,Guarnera:1996svb}. Hybrid approaches have coupled hydrodynamics to mean field dynamics, e.g. for scalar fields related to chiral symmetry, and in those approaches gradient terms come into play for the fields \cite{Paech:2003fe,Paech:2005cx,Nahrgang:2011mg}. Larger gradient terms more strongly disfavor sharp changes in the order parameter, and result in larger surface energies, longer correlation lengths, larger nucleation barriers, and larger damping for the growth of short wavelength unstable modes in hydrodynamically unstable regions.

Our goal here is to incorporate these effects into a dynamic theory, hydrodynamics, and to determine how to consistently apply gradient modifications to all thermodynamic quantities: temperature, entropy density, chemical potentials and the stress-energy tensor. Hydrodynamics plays a critical role in the modeling of relativistic heavy-ion collisions \cite{Kolb:2003dz}. Fluctuations and related correlations have been measured at the Relativistic Heavy-Ion Collider (RHIC) and for heavy-ion collisions at the LHC \cite{Adams:2005aw,Sharma:2009zt,Adams:2006sg,Abelev:2014ckr,Abelev:2012pv,Adamczyk:2013up,Abelev:2009ai,Abelev:2008jg,Tarnowsky:2011vk,Adamczyk:2013hsi}, where they play an important role in understanding the phase structure and chemistry of the highly excited matter created in these collisions. For example, correlations driven by local charge conservation \cite{Pratt:2012dz,Pratt:2015jsa} appear especially promising for making comparison to charge fluctuations extracted from lattice gauge theory \cite{Borsanyi:2011sw,Bellwied:2015lba} and \cite{Borsanyi:2014ewa}. 

Correlations of transverse energy and momentum provide insight into thermodynamic properties, diffusion and initial state fluctuations such as those from jets \cite{CaronHuot:2009iq,Akamatsu:2016llw,Pang:2009zm,Gavin:2006xd,Gavin:2012if,Gavin:2016hmv,Gavin:2016jfw,Gavin:2016nir,Pratt:2010zn,Pratt:2016lol}. Correlations from charge conservation also play an important role for understanding the background for measurements related to the chiral magnetic effect \cite{starcme,Schlichting:2010qia,Pratt:2010zn}.  As the field works to consider the growth of thermal fluctuations, especially those related to phase structure as near a critical point, gradient terms are important in modeling both the size and spread of correlations. 

In hydrodynamics the evolution is driven by local conservation of the stress energy tensor,
\begin{eqnarray}
\partial_\mu T^{\mu\nu}=0.
\end{eqnarray}
In the fluid rest frame $T^{00}$ is the energy density $\epsilon$, $T^{0i}$ are the momentum densities, and $T_{ij}$ represents the pressure tensor and is typically a function of $\epsilon$ and the charge density $\rho$. For ideal hydrodynamics $T_{ij}=P\delta_{ij}$ in the fluid's rest frame (we use Roman indices to indicate the spatial indices), and the pressure $P$ is a function of $\epsilon$ and $\rho$. Non-diagonal terms then appear when the tensor is viewed in a frame where the fluid moves with a velocity $v$. Once the initial conditions are set, the ensuing evolution is determined by $P(\epsilon,\rho)$. The entropy within a co-moving cell is fixed. Entropy and temperature are key concepts of thermal systems, but do not necessarily appear in the equations because the hydrodynamic equations of motion involve only $\epsilon$, $\rho$ and the fluid velocity $\bm{v}$.

Here, we show how gradient terms can be incorporated into the equations of hydrodynamics. Often, mean-field theories assume a globally fixed quantity, usually the temperature. In that case the system adjusts the density to minimize the Helmholtz free energy, $F=E-TS$. If the resulting density profile is stable, e.g. at a phase boundary, the chemical potentials should be constant. Otherwise, the entropy could be increased by moving charge from a region of higher chemical potential to a region of lower chemical potential. Further, an equilibrated profile must also have vanishing hydrodynamic acceleration, $\partial_iT_{ij}=0$. This requires adding gradient modifications to both the chemical potential and stress-energy tensor. The principal goal of this paper is to develop a formalism where all the gradient modifications appear in a manner which is consistent with local entropy conservation. A second consistency check of the formalism is that in a density profile where the temperature and chemical potentials remain constant, the stress-energy tensor should also be free of acceleration. 

General theoretical derivations are presented in the next section, and a discussion of alternate formulations is presented in Sec. \ref{sec:alternate}. After assuming a form for the entropy density with gradient modifications, corrections for other quantities are then uniquely determined. We find that all the dependent quantities, $\mu$, $T$ and $T_{ij}$, are uniquely altered to maintain the consistencies described above. 
In Sec.s \ref{sec:liquidgas} and \ref{sec:thermalflucs}, two static examples are investigated, the phase boundary of the liquid gas phase transition, and the form for density-density correlations. In each case, the effect of the gradient terms are illustrated. In Sec. \ref{sec:noise}, a consistent form for thermal noise in the current-current correlation function, \cite{Kapusta:2014dja,Kapusta:2012sd,Young:2014pka,Ling:2013ksb,Pratt:2016lol}, is presented that accounts for the gradient terms. 

\section{Implementing Gradient Terms into Hydrodynamics}
\label{sec:theory}

For hydrodynamics, the natural quantities to describe the system are the energy density $\epsilon$, the charge density $\rho$, and the collective velocity $\bm{v}$. For our purposes, we consider the Eckart frame, where the velocity defines the movement of charge, i.e. when $\bm{v}=0$ the charge current vanishes, and $\epsilon=T^{00}$ in that frame, while $\rho=j^0$, the component of the four-current in that same frame. For fixed $\epsilon$ and $\rho$, the natural thermodynamic quantity to consider is the entropy density, which for a uniform system would have a form $\bar{s}(\epsilon,\rho)$. Given that hydrodynamic cells have fixed energy and charge, the system adjusts toward maximizing entropy. Here, gradient modifications are added to the expression for the entropy density with the form,
\begin{eqnarray}
\label{eq:Sdef}
s&=&\bar{s}\left(\epsilon_\kappa,\rho\right),\\
\nonumber
\epsilon_\kappa&=&\epsilon+\frac{\kappa}{2}\rho\nabla^2\rho.
\end{eqnarray}
Here, all quantities with the bar, e.g. $\bar{s}$, refer to the value one would have with uniform density. The form is motivated by considering the potential energy of a particle a position $\delta\bm{r}$ relative to its neighbors. The contribution from an attractive potential, $v(\bm{r}-\bm{r}')$, with other charges is
\begin{eqnarray}
V&=&\frac{1}{2}\int d^3rd^3r'~\rho(\bm{r})v({\bm r}-\bm{r}')\rho(\bm{r}')\\
\nonumber
\nonumber&=&\frac{1}{2}\int d^3r ~\rho(\bm{r})\int d^3\delta r~v(\bm{\delta r})\left[\rho(\bm{r})+\frac{(\delta r)^2}{6}\nabla^2\rho|_{\bm{r}=0}\right],\\
\nonumber
\Delta V&=&-\frac{1}{2}\int d^3r~\kappa\rho\nabla^2\rho,\\
\kappa&=&-\frac{1}{6}\int d^3r~r^2v(r).
\end{eqnarray}
Here, $\Delta V$ is the change in energy due to the fact that neighbors are not at a uniform density. These assumptions ignore correlations between the particles, which would suggest a density or temperature dependence for $\kappa$. Adding potential energy to a neighborhood of particles should not strongly affect the entropy, so the entropy should mainly be a function of the energy one would have without this correction, $\epsilon_\kappa=\epsilon+\kappa\rho\nabla^2\rho/2$, hence the form for Eq. (\ref{eq:Sdef}).

A more general form for Eq. (\ref{eq:Sdef}) would allow $\kappa$ to depend on $\epsilon$ and $\rho$, but for the sake of simplicity, those dependencies are neglected here given the phenomenological nature of this treatment where $\kappa$ will be treated as an adjustable parameter.  We also neglect the possibility of additional terms such as
\begin{eqnarray}
\label{eq:Sdefcomplicated}
\epsilon_\kappa=\epsilon+\frac{1}{2}\kappa_{\rho\rho}\rho\nabla^2\rho+\frac{1}{2}\kappa_{\epsilon\epsilon}\epsilon\nabla^2\epsilon+\frac{1}{2}\kappa_{\epsilon\rho}(\epsilon\nabla^2\rho+\rho\nabla^2\epsilon).
\end{eqnarray}
For a charge-neutral system, there are only terms with gradients of $\epsilon$, and a complete delineation of second-order terms can be found in \cite{Romatschke:2009kr}. An exhaustive list of such terms is not the goal of this study, but rather to consider the simplest expression that captures the physics necessary for studying phenomena related to phase transitions, especially critical fluctuations, phase bounaries and phase separation. For a quark-gluon plasma with zero net charge one could envision an attractive interaction between induced color dipoles. If such effects were strong, they might help create the conditions for a first-order phase transition and there would be a surface energy at the interface \cite{Csernai:1992tj,Csernai:2002bu}. There is no such transition, thus such effects are probably small and we consider only the simple form of Eq. (\ref{eq:Sdef}) and neglect terms with gradients of $\epsilon$ for the remainder of this paper.

One can quickly calculate corrections to the temperature and chemical potential from Eq. (\ref{eq:Sdef}) by considering changes to the total entropy due to changes in the energy density and charge density. First, we consider the conditions for the entropy $S$ being maximized relative to small corrections in $\delta\epsilon$.
\begin{eqnarray}
S&=&\int d^3r~\bar{s}(\epsilon+\kappa\rho\nabla^2\rho/2,\rho),\\
\nonumber
\delta S&=&\int d^3r~\beta \delta\epsilon,\\
\nonumber
\beta&=&\left.\frac{\partial\bar{s}}{\partial\epsilon}\right|_{\epsilon=\epsilon_\kappa}=\bar{\beta}(\epsilon_\kappa,\rho).
\end{eqnarray}
Here, $\beta=1/T$ is the inverse temperature. This differs from the inverse temperature calculated for a uniform system, $\bar{\beta}(\epsilon,\rho)$.
Similarly, one can find the alterations to the chemical potential by considering small changes in charge density,
\begin{eqnarray}
\delta S&=&\int d^3r~\left[\bar{\beta}(\epsilon_\kappa,\rho)\kappa(\delta\rho)\nabla^2\rho/2
+\bar{\beta}(\epsilon_\kappa,\rho)\kappa\rho\nabla^2(\delta\rho)/2
+\bar{\alpha}(\epsilon_\kappa,\rho)(\delta\rho)\right].
\end{eqnarray}
Here, $\alpha$ is related to the chemical potential by the relation $\alpha=-\beta\mu$. The function $\bar{\alpha}(\epsilon_\kappa,\rho)$ represents the chemical potential for the case of uniform charge and energy density. After integrating by parts, one finds the chemical potential,
\begin{eqnarray}
\label{eq:alpharesult}
\delta S&=&\int d^3r~\delta\rho\left[\frac{\bar{\beta}(\epsilon_\kappa,\rho)\kappa}{2}\nabla^2\rho+\frac{\kappa}{2}\nabla^2(\bar{\beta}(\epsilon_\kappa,\rho)\rho)+\bar{\alpha}(\epsilon_\kappa,\rho)\right],\\
\nonumber
\alpha&=&\bar{\alpha}+\frac{\bar{\beta}\kappa}{2}\nabla^2\rho+\frac{\kappa}{2}\nabla^2(\bar{\beta}\rho),\\
\nonumber
\bar{\alpha}(\epsilon_\kappa,\rho)&=&\left.\frac{\partial s}{\partial\rho}\right|_{\epsilon=\epsilon_\kappa}.
\end{eqnarray}
In this expression, and in the following derivations, the quantities with bars, e.g. $\bar{\alpha}$ and $\bar{\beta}$ are evaluated at $\epsilon_\kappa$ and $\rho$, unless explicitly shown to be evaluated at $\epsilon$.

The next, and more difficult, goal is to discern how the gradient terms affect the stress-energy tensor. To do this we consider the change in entropy density during an expansion, i.e. in the presence of a velocity gradient. In that case, entropy conservation gives
\begin{eqnarray}
\label{eq:currentcons}
D_ts&=&-s\bm{\nabla}\cdot\bm{v},
\end{eqnarray}
where here it is implied that the entropy current vanishes at $\bm{v}=0$ and is $s\bm{v}$ for small $\bm{v}$. Here, $D_t=\partial_t+\bm{v}\cdot\bm{\nabla}$ is the co-moving derivative, i.e. it is $\partial_t$ when in the frame where $\bm{v}=0$. 

Following through with the calculation for entropy conservation,
\begin{eqnarray}
\label{eq:Dts0}
D_ts&=&\bar{\beta}\left[D_t\epsilon+D_t\left(\frac{\kappa}{2}\rho\nabla^2\rho\right)\right]
+\bar{\alpha}D_t\rho.
\end{eqnarray}
Next, we apply conservation of energy and charge,
\begin{eqnarray}
\label{eq:Dtrho}
D_t\rho&=&-\rho\bm{\nabla}\cdot\bm{v},\\
\nonumber
D_t\epsilon&=&-\epsilon\bm{\nabla}\cdot\bm{v}-T_{ij}\partial_iv_j-\partial_iM_i,
\end{eqnarray}
where $\bm{M}$ is the momentum density, and represents the components of the stress-energy tensor $M_i=T^{0i}$. Equations (\ref{eq:currentcons}) and (\ref{eq:Dtrho}) are based on the physical picture where the charge density and entropy density move together, whereas the energy might have some additional current $\bm{M}$, even in the frame where $\bm{v}=0$. This is not the case in ideal hydrodynamics, where there is no heat conduction or charge diffusion. However, here there are terms from non-zero-range interactions. Because these interactions extend over a non-zero distance between two charges, it is somewhat ambiguous to assign the position of this portion of the potential energy. Movement of charge outside a given small co-moving fluid cell can affect the energy within that cell, and therefore the energy might move, represent to the cell even though the thermal motion of matter within the cells is unchanged. Such currents should only exist in the presence of velocity gradients, and should be proportional to $\kappa$. In contrast, the position of charge is well defined, and because the entropy should not be affected by changes in the potential energy, the entropy current should also vanish in the frame where the charge current vanishes. Thus, there are no additional terms on the right-hand sides of the expressions for $D_ts$ and $D_t\rho$.

Expanding Eq. (\ref{eq:Dts0}),
\begin{eqnarray}
D_ts&=&-\bar{\beta}T_{ij}\partial_iv_j-\left[\bar{\alpha}\rho+\bar{\beta}\epsilon\right]\bm{\nabla}\cdot\bm{v}
+\bar{\beta}D_t\left(\frac{\kappa}{2}\rho\nabla^2\rho\right)-\bar{\beta}\bm{\nabla}\cdot\bm{M}.
\end{eqnarray}
Next, one can commute $D_t$ with $\nabla^2$,
\begin{eqnarray}
\left[D_t,\nabla^2\right]&=&-2(\partial_iv_j)\partial_j\partial_i-(\nabla^2v_i)\partial_i,
\end{eqnarray}
and find
\begin{eqnarray}
\label{eq:DtSa}
D_ts&=&-\bar{\beta}T_{ij}\partial_iv_j-\left[\bar{\alpha}\rho+\bar{\beta}\epsilon\right]\bm{\nabla}\cdot\bm{v}-\bar{\beta}\bm{\nabla}\cdot\bm{M}\\
\nonumber
&&-\bar{\beta}\kappa\left\{\rho(\nabla^2\rho)(\bm{\nabla}\cdot\bm{v})
+\rho(\partial_i\partial_j\rho)\partial_iv_j
+\rho(\partial_i\rho)(\partial_i\bm{\nabla}\cdot\bm{v})
+\frac{1}{2}\rho(\partial_i\rho)(\nabla^2v_i)
+\frac{1}{2}\rho^2(\nabla^2\bm{\nabla}\cdot\bm{v})
\right\}.
\end{eqnarray}
Again, $\bar{\beta}$ and $\bar{\alpha}$ are implicitly evaluated at $\epsilon_\kappa$ and $\rho$. The last three terms include higher derivatives of the velocity. Because our goal is to find $T_{ij}$ and because derivatives of the form $\bar{\beta}\bm{\nabla}\cdot(\cdots)$ can be absorbed into the expression for $\bm{M}$, we rewrite the last three terms so that they appear as a combination of total derivatives or only include first-order derivatives of the velocity. For example the last term can be written as
\begin{eqnarray}
-\frac{1}{2}\kappa\rho^2(\nabla^2\bm{\nabla}\cdot\bm{v})
&=&-A\partial_i\left\{\frac{\kappa}{2}\rho^2(\partial_i\bm{\nabla}\cdot\bm{v})\right\}
-(1-A)\partial_i\left\{\frac{\kappa}{2}\rho^2(\nabla^2v_i)\right\}\\
\nonumber
&&+A(\partial_i\bm{\nabla}\cdot\bm{v})\partial_i\left\{\frac{\kappa}{2}\rho^2\right\}
+(1-A)(\nabla^2v_i)\partial_i\left\{\frac{\kappa}{2}\rho^2\right\}.
\end{eqnarray}
Here, $A$ is an arbitrary constant. The first two terms, which are total derivatives, can be canceled in the equation for entropy conservation, Eq. (\ref{eq:DtSa}), by an equivalent term in $\bm{M}$. The latter two terms include second-order derivatives in the velocity rather than third order. Similar manipulations can then reduce these terms with second-order derivatives of the velocity to terms with first-order derivatives, along with other total derivatives. This yields
\begin{eqnarray}
\label{eq:D_tsresult}
\bar{T}D_ts&=&-T_{ij}\partial_iv_j-(\bar{\mu}\rho+\epsilon)\bm{\nabla}\cdot\bm{v}
\\ \nonumber &&
+\frac{\kappa}{2}\partial_i\left[\rho\partial_j\rho-(1-A)(\partial_j(\rho^2))\right]\partial_iv_j
+B(\partial_i(\rho\partial_i\rho))\bm{\nabla}\cdot\bm{v}
-\frac{\kappa}{2}AC(\nabla^2(\rho^2))\bm{\nabla}\cdot\bm{v}
\\ \nonumber &&
+(1-B)\kappa(\partial_i(\rho\partial_j\rho))\partial_jv_i
-A(1-C)\frac{\kappa}{2}(\partial_i\partial_j(\rho^2))\partial_jv_i
\\ \nonumber &&
-\kappa\rho(\nabla^2\rho)\bm{\nabla}\cdot\bm{v}
-\kappa\rho(\partial_i\partial_j\rho)\partial_iv_j.
\end{eqnarray}
Terms of form of total derivatives, $\bm{\nabla}(\cdots)$, were eliminated by defining the momentum density as
\begin{eqnarray}
M_i&=&-A\frac{\kappa}{2}\rho^2\partial_i\bm{\nabla}\cdot\bm{v}
-(1-A)\frac{\kappa}{2}\rho^2\nabla^2v_i
\\ \nonumber && 
-\frac{\kappa}{2}\rho(\partial_j\rho)\partial_iv_j
+(1-A)\frac{\kappa}{2}(\partial_j(\rho^2))\partial_iv_j
-B\kappa\rho(\partial_i\rho)\bm{\nabla}\cdot\bm{v}
+AC\frac{\kappa}{2}(\partial_i(\rho^2))\bm{\nabla}\cdot\bm{v}
\\ \nonumber && 
-(1-B)\rho(\partial_j\rho)\partial_iv_j-A(1-C)\frac{\kappa}{2}(\partial_j(\rho^2))\partial_jv_i
\end{eqnarray}
Here, $A$, $B$ and $C$ are all arbitrary constants.  Only terms with first-order derivatives in the velocity remain in Eq. (\ref{eq:D_tsresult}), so $T_{ij}$ can be determined by inspection.

The additional momentum density $\bm{M}$ refers only to that part of that momentum density that exists in the frame $\bm{v}=0$. Further, $\bm{M}$ has no terms proportional to $\bm{v}$. Those terms are generated by boosting the entire stress-energy tensor, which generates a contribution $\epsilon\delta v_i+T_{ij}\delta v_j$. If there were terms in $\bm{M}$ linear in $\delta v$, they would violate the expected behavior of the stress-energy tensor under boosts. Another class of transformations that leaves the relative density profile, and thus the gradient modification to the energy density unchanged, is rotation, $\delta\bm{v}=\bm{r}\times\delta\bm{\omega}$. Rotations should also not generate  momentum density from the gradient term, aside from contribution from rotating the stress-energy tensor. To satisfy this constraint the expression for $\bm{M}$ must avoid terms that depend on $\omega_i=\epsilon_{ijk}\partial_jv_k$. Thus, all velocity gradients must appear either as the symmetric combination, $\partial_iv_j+\partial_jv_i$, or as $\bm{\nabla}\cdot\bm{v}$. This requires
\begin{eqnarray}
A&=&1, B=\frac{1}{2}, C=1.
\end{eqnarray}
The momentum density is then
\begin{eqnarray}
\label{eq:Jsol}
M_i&=&-\frac{\kappa}{2}\rho(\partial_j\rho)(\partial_iv_j+\partial_jv_i)
-\frac{\kappa}{2}\rho^2\partial_i\bm{\nabla}\cdot\bm{v}
+\frac{\kappa}{2}\rho(\partial_i\rho)\bm{\nabla}\cdot\bm{v}.
\end{eqnarray}
For an expanding or contracting system, i.e. one with velocity gradients, the potential energy of a constituent charge changes due to the changing relative positions of its neighbors. Energy moves from cell to cell, but the net energy is unchanged.  

The equation for entropy conservation, Eq. (\ref{eq:currentcons}), then becomes
\begin{eqnarray}
&&-\bar{\beta}T_{ij}\partial_iv_j-\bar{\alpha}\rho\bm{\nabla}\cdot\bm{v}-\bar{\beta}\epsilon\bm{\nabla}\cdot\bm{v}
\\ \nonumber &&
+\frac{\kappa\bar{\beta}}{2}(\partial_i(\rho\partial_j\rho))(\partial_iv_j+\partial_jv_i)
+\frac{\kappa\bar{\beta}}{2}(\partial_i(\rho\partial_i\rho))\bm{\nabla}\cdot\bm{v}
-\frac{\kappa\bar{\beta}}{2}(\nabla^2(\rho^2))\bm{\nabla}\cdot\bm{v}
\\ \nonumber &&
-\bar{\beta}\kappa\rho(\nabla^2\rho)\bm{\nabla}\cdot\bm{v}
-\bar{\beta}\kappa\rho(\partial_i\partial_j\rho)\partial_iv_j
\\ \nonumber
&=&-s\bm{\nabla}\cdot\bm{v}
\end{eqnarray}
Using the fact that $s=\bar{s}$ and $\bar{s}-\bar{\alpha}\rho-\bar{\beta}\epsilon_\kappa=\bar{P}$, where $\bar{P}$ is also evaluated at $\epsilon_\kappa$, one finds 
\begin{eqnarray}
\label{eq:Tijresult}
\bar{\beta}T_{ij}\partial_iv_j&=&\bar{\beta}\bar{P}\bm{\nabla}\cdot\bm{v}
-\frac{\bar{\beta}\kappa}{2}\rho(\nabla^2\rho)\bm{\nabla}\cdot\bm{v}
+\frac{\kappa\bar{\beta}}{2}(\partial_i(\rho\partial_j\rho))(\partial_iv_j+\partial_jv_i)
\\ \nonumber &&
+\frac{\kappa\bar{\beta}}{2}(\partial_i(\rho\partial_i\rho))\bm{\nabla}\cdot\bm{v}
-\frac{\kappa\bar{\beta}}{2}(\nabla^2(\rho^2))\bm{\nabla}\cdot\bm{v}
-\bar{\beta}\kappa\rho(\partial_i\partial_j\rho)\partial_iv_j,\\
\nonumber
T_{ij}&=&
\delta_{ij}\left\{\bar{P}-\frac{\kappa}{2}\rho\nabla^2\rho
+\frac{\kappa}{2}(\partial_i(\rho\partial_i\rho))
-\frac{\kappa}{2}(\nabla^2(\rho^2))\right\}
\\ \nonumber &&
+\frac{\kappa}{2}\partial_i\left(\rho\partial_j\rho\right)
+\frac{\kappa}{2}\partial_j\left(\rho\partial_i\rho\right)
-\kappa\rho(\partial_i\partial_j\rho)\\
\nonumber
&=&\bar{P}\delta_{ij}
-\kappa\left[\rho\nabla^2\rho+\frac{1}{2}(\bm{\nabla}\rho)^2\right]\delta_{ij}
+\kappa(\partial_i\rho)\partial_j\rho.
\end{eqnarray}
Thus, $T_{ij}$ is not purely proportional to $\delta_{ij}$, and one cannot express the change to the stress-energy tensor by only altering the pressure.

In Sec. \ref{sec:liquidgas} we consider the example of a surface profile between two phases at equilibrium. For entropy to be maximized both $\beta$ and $\alpha$ must be uniform. Also, if entropy is maximized there should be no hydrodynamic acceleration. Thus, if the temperature is uniform and if $\bm{v}=0$, $\partial_iT_{ij}$, must vanish if $\partial_i\mu=0$. Otherwise, charge or energy could be moved from one point to another while increasing entropy. Using the expression for the stress-energy tensor in Eq. (\ref{eq:Tijresult}), at equilibrium
\begin{eqnarray}
\label{eq:dTij0}
\partial_iT_{ij}^{\rm(equil)}&=&0\\
\nonumber
&=&\partial_j\bar{P}-\kappa\rho\partial_j\nabla^2\rho.
\end{eqnarray}
For uniform temperature, the pressure gradient can be related to the chemical potential gradient
\begin{eqnarray}
\partial_i\bar{P}&=&\frac{\partial\bar{P}}{\partial\bar{\mu}}\partial_i\bar{\mu}=\rho\partial_i\bar{\mu},
\end{eqnarray}
and from Eq.s (\ref{eq:alpharesult}) and (\ref{eq:dTij0}),
\begin{eqnarray}
\label{eq:alphaconsistency}
\partial_iT_{ij}^{\rm(equil)}&=&\rho\partial_j\left(\bar{\mu}-\kappa\rho\nabla^2\rho\right)=\rho\partial_j\mu.
\end{eqnarray}
Thus, if a profile has a uniform chemical potential and temperature, even if the density is non-uniform, hydrodynamic acceleration vanishes. Because uniform $\mu$ and $T$ should be sufficient to maximize entropy, the system should also be free of acceleration and Eq. (\ref{eq:alphaconsistency}) represents a test of the consistency of the formalism.

Finally, the gradient terms also affect the expressions for diffusion, or equivalently, the conductivity. Diffusion is essential to include if a treatment is to reproduce equilibrium quantities, e.g. for a phase boundary. At equilibrium, one must satisfy the relations $\partial_iT_{ij}=0$, $\partial_i\beta=0$ and $\partial_i\mu=0$. Hydrodynamically, a system can expand (with damping) until the acceleration disappears, but this would not necessarily lead to a uniform temperature and chemical potential, because hydrodynamic equations would maintain a uniform entropy to baryon ratio. In contrast, the entropy per baryon varies significantly across a phase boundary. By allowing charge to diffuse between neighboring hydrodynamic cells, the system can equilibrate and eliminate gradients of all three quantities. Thus, if a model is designed to study dynamics related to either reaching equilibrium, or moving toward equilibrium, both hydrodynamic motion and diffusion should be included, and in a manner consistent with thermodynamics.

Compared to the frame where the entropy current is fixed, diffusion moves the charge density via Fick's law.
\begin{eqnarray}
\bm{J}^{\rm(D)}&=&\sigma T\bm{\nabla}\alpha=-\sigma T\bm{\nabla}(\mu/T).
\end{eqnarray}
Here, aside from powers of the charge, $\sigma$ is the conductivity. This can be related to the diffusion constant,
\begin{eqnarray}
\bm{\nabla}\rho&=&\frac{\partial\rho}{\partial\bar{\alpha}}\bm{\nabla}\bar{\alpha}\\
\nonumber
&=&-\chi\bm{\nabla}\bar{\alpha},
\end{eqnarray}
where $\chi$ is the susceptiblity, or charge fluctuation,
\begin{eqnarray}
\chi&=&\left.T\frac{\partial\rho}{\partial\mu}\right|_T=\frac{1}{V}\left\langle (Q-\langle Q\rangle)^2\right\rangle.
\end{eqnarray}
The diffusion constant $D$ is then related to the conductivity through the relation
\begin{eqnarray}
\bm{j}&=&-D\bm{\nabla}\rho,\\
\nonumber
D&=&\frac{\sigma T}{\chi}.
\end{eqnarray}
From the gradient modification to the chemical potential, Eq. (\ref{eq:alpharesult}),
\begin{eqnarray}
\label{eq:Fick}
\vec{J}^{\rm(D)}&=&\sigma T\bm{\nabla}\left(\bar{\alpha}+\frac{\bar{\beta}\kappa}{2}\nabla^2\rho+\frac{\kappa}{2}\nabla^2(\bar{\beta}\rho)\right)\\
\nonumber
&=&-\frac{\sigma T}{\chi}\bm{\nabla}\rho+\frac{\sigma \kappa T}{2}\bm{\nabla}\left[\bar{\beta}\nabla^2\rho+\nabla^2(\bar{\beta}\rho)\right]\\
\nonumber
&=&-D\bm{\nabla}\rho+\frac{\kappa\chi D}{2}\bm{\nabla}\left[\bar{\beta}\nabla^2\rho+\nabla^2(\bar{\beta}\rho)\right].
\end{eqnarray}
Thus, there are higher-order gradient modifications to Fick's law. 

Summarizing the results of this section, gradient terms affect every thermodynamic quantity,
\begin{eqnarray}
\label{eq:summary}
s&=&\bar{s}(\epsilon_\kappa,\rho),\\
\nonumber
\beta&=&\bar{\beta}(\epsilon_\kappa,\rho),\\
\nonumber
\alpha&=&\bar{\alpha}(\epsilon_\kappa,\rho)+\frac{\bar{\beta}\kappa}{2}\nabla^2\rho+\frac{\kappa}{2}\nabla^2(\bar{\beta}\rho),\\
\nonumber
M_i&=&-\frac{\kappa}{2}\rho(\partial_j\rho)(\partial_iv_j+\partial_jv_i)
-\frac{\kappa}{2}\rho^2\partial_i\bm{\nabla}\cdot\bm{v}
+\frac{\kappa}{2}\rho(\partial_i\rho)\bm{\nabla}\cdot\bm{v}.,\\
\nonumber
T_{ij}&=&\bar{P}\delta_{ij}
-\kappa\left[\rho\nabla^2\rho+\frac{1}{2}(\bm{\nabla}\rho)^2\right]\delta_{ij}
+\kappa(\partial_i\rho)(\partial_j\rho),\\
\nonumber
\bm{J}^{\rm(D)}&=&-D\bm{\nabla}\rho+\frac{\kappa\chi D}{2}\bm{\nabla}\left[\bar{\beta}\nabla^2\rho+\nabla^2(\bar{\beta}\rho)\right].
\end{eqnarray}
Each of the quantities with bars are those for uniform energy and charge density, but are evaluated at an energy density $\epsilon_\kappa=\epsilon+\kappa\rho\nabla^2\rho/2$, and at a charge density $\rho$. Once the entropy density is defined in the first line of Eq. (\ref{eq:summary}), all the other relations are uniquely determined. 

Because the ansatz for the entropy density in the first line of Eq.s (\ref{eq:summary}) has $\kappa$ being independent of $\epsilon$ and $\rho$, and because terms involving derivatives of the energy density are ignored, these relations are far from completely general. However, the form is justifiable in many circumstances, especially given the phenomenological nature of most studies based on Landau field theory. The forms are consistent with thermodynamics even in dynamical situations where the temperature, stress-energy tensor and chemical potentials vary with both position and time.

\section{Alternate Forms}
\label{sec:alternate}

The equations from the previous section, summarized in Eq. (\ref{eq:summary}), all derived from Eq. (\ref{eq:Sdef}). If this original ansatz is changed, then the resulting equations also change. For example, in the studies \cite{Steinheimer:2013xxa,Steinheimer:2013gla,Steinheimer:2012gc,Randrup:2010ax}, a choice is made for the pressure, $P=\bar{P}+\kappa\rho\nabla^2\rho$, which gives a diagonal form to the stress-energy tensor in the fluid frame, and is much simpler  than what is seen in Eq. (\ref{eq:summary}). However, it is difficult to reconcile this form with the form for the chemical potential for a system at uniform temperature, $\mu=\bar{\mu}-\kappa\nabla^2\rho$. Even at uniform temperature, this form fails to provide the consistency demonstrated in Eq. (\ref{eq:alphaconsistency}).  To understand these difficulties, one can consider a phase boundary at equilibrium, or for that matter fluctuations at equilibrium. The temperature and chemical potentials must be constant across the profile. Otherwise, charge or energy could be transferred in such a way as to increase the net entropy. The system must also be simultaneously stable to hydrodynamic acceleration, $\partial_iT_{ij}=0$. The gradient modifications to the stress-energy tensor at uniform temperature might have a general form,
\begin{eqnarray}
T_{ij}=\bar{P}\delta_{ij}+A\delta_{ij}\rho\nabla^2\rho +B\delta_{ij}(\bm{\nabla}\rho)^2 +C(\partial_i\rho)\partial_j\rho +D\rho\partial_i\partial_j\rho.
\end{eqnarray}
Given that $\bm{\nabla}\bar{P}=(\partial P/\partial\mu)\bm{\nabla}\bar{\mu}=\rho\bm{\nabla}\bar{\mu}$, if $\partial_iT_{ij}$ is to vanish any time that $\partial_j\mu$ vanishes, one must have
\begin{eqnarray}
\partial_iT_{ij}&=&\rho\partial_j\mu,\\
\nonumber
\partial_i\left\{A\delta_{ij}\rho\nabla^2\rho +B\delta_{ij}(\bm{\nabla}\rho)^2 +C(\partial_i\rho)\partial_j\rho +D\rho\partial_i\partial_j\rho\right\}
&=&-\kappa\rho\partial_j\nabla^2\rho,\\
\nonumber
\rho(\partial_j\nabla^2\rho)(A+C)+(\partial_i\rho)(\partial_i\partial_j\rho)(2B+C+D)+(\partial_j\rho)(\nabla^2\rho)(A+D)&=&-\kappa\rho(\partial_j\nabla^2\rho).
\end{eqnarray}
This determines three of the four parameters. In terms of $C$,
\begin{eqnarray}
A=-\kappa-C,~~~B=-\frac{\kappa+2C}{2},~~~D=\kappa+C.
\end{eqnarray}
This constraint was indeed satisfied by the relations in Eq. (\ref{eq:summary}). In that derivation $C$ was zero in order to enforce proper behavior of the entropy current under rotation. From this exercise, it is clear that the stress-energy tensor needs to have off-diagonal terms, because both $C$ and $D$ cannot be set to zero. This suggests that it would be problematic to build a consistent set of gradient modifications to all thermodynamic quantities if one wishes to use a simple form for the stress-energy tensor with no gradient modifications to off-diagonal terms.

\section{The Liquid-Gas Phase Interface}
\label{sec:liquidgas}

Here, we illustrate the effect of gradient modifications by considering the static density profile between liquid and gas phases at equilibrium, similarly as was done in \cite{Ravenhall:1984ss}. We consider the system to be at a uniform fixed temperature, then solve for the density profile by requiring the chemical potential to be constant, or equivalently requiring that diffusion vanishes. After finding the density profile, we compare the stress-energy tensor profile, temperature and chemical potential to the values one would have ignoring gradient terms, $\bar{T}_{ij}=\bar{P}\delta_{ij}$, $\bar{\beta}$ and $\bar{\alpha}$, evaluated at the energy density $\epsilon$ instead of $\epsilon_\kappa$. 

We assume the Van der Waals equation of state, 
\begin{eqnarray}
\bar{P}&=&\frac{\rho\bar{T}(\epsilon_\kappa,\rho)}{1-\rho/\rho_s}-a\rho^2.
\end{eqnarray}
The critical temperature for this equation of state is $T_c=(8/27)a\rho_s$. As stated before, $\bar{P}$ and $\bar{T}$ are the pressure and temperature for a uniform system. At equilibrium, $\bar{T}$ can be treated as a constant. The chemical potential, $\mu=\bar{\mu}-\kappa\rho\nabla^2\rho/2$, is then also constant. For the fixed temperature, $\bar{T}$, one can solve for the liquid and gas coexistence densities by requiring
\begin{eqnarray}
\bar{\mu}(\bar{T},\rho_L)&=&\bar{\mu}(\bar{T},\rho_G),\\
\nonumber
\bar{P}(\bar{T},\rho_L)&=&\bar{P}(\bar{T},\rho_G),\\
\nonumber
\bar{\mu}&=&\frac{\partial(\bar{P}-\bar{\mu}\rho)}{\partial\rho}\\
\nonumber
&=&-2a\rho+\frac{T}{1-\rho/\rho_s}-T\ln((\rho_s/\rho)-1).
\end{eqnarray}
The last relation was derived via the Maxwell relation, $\rho\partial\mu/\partial\rho=\partial P/\partial\rho$. Minimizing the entropy per surface area, relative to moving particles to the heat bath, gives the condition
\begin{eqnarray}
T\delta S/A&=&0\\
\nonumber
&=&\delta\int dx~\left[P_0-P+(\mu-\mu_0)\rho+\frac{\kappa}{2}(d\rho/dx)^2\right]\\
&=&\int_{\rho_G}^{\rho_L}d\rho~\delta\left[\frac{P_0-P+(\mu-\mu_0)\rho+(\kappa/2)(d\rho/dx)^2}{d\rho/dx}\right],
\end{eqnarray}
where the phase interface is parallel to the $x$ axis. Here, $P_0=P_G=P_L$ and $\mu_0=\mu_G=\mu_L$ are the coexistence values. The density gradient is then
\begin{eqnarray}
\frac{d\rho}{dx}&=&\sqrt{2[P_0-P(\rho)+(\mu(\rho)-\mu_0)\rho]/\kappa}.
\end{eqnarray}
This expression can be integrated numerically to give $x(\rho)$ and thus determine the density profile.

\begin{figure}
\centerline{\includegraphics[width=0.45\textwidth]{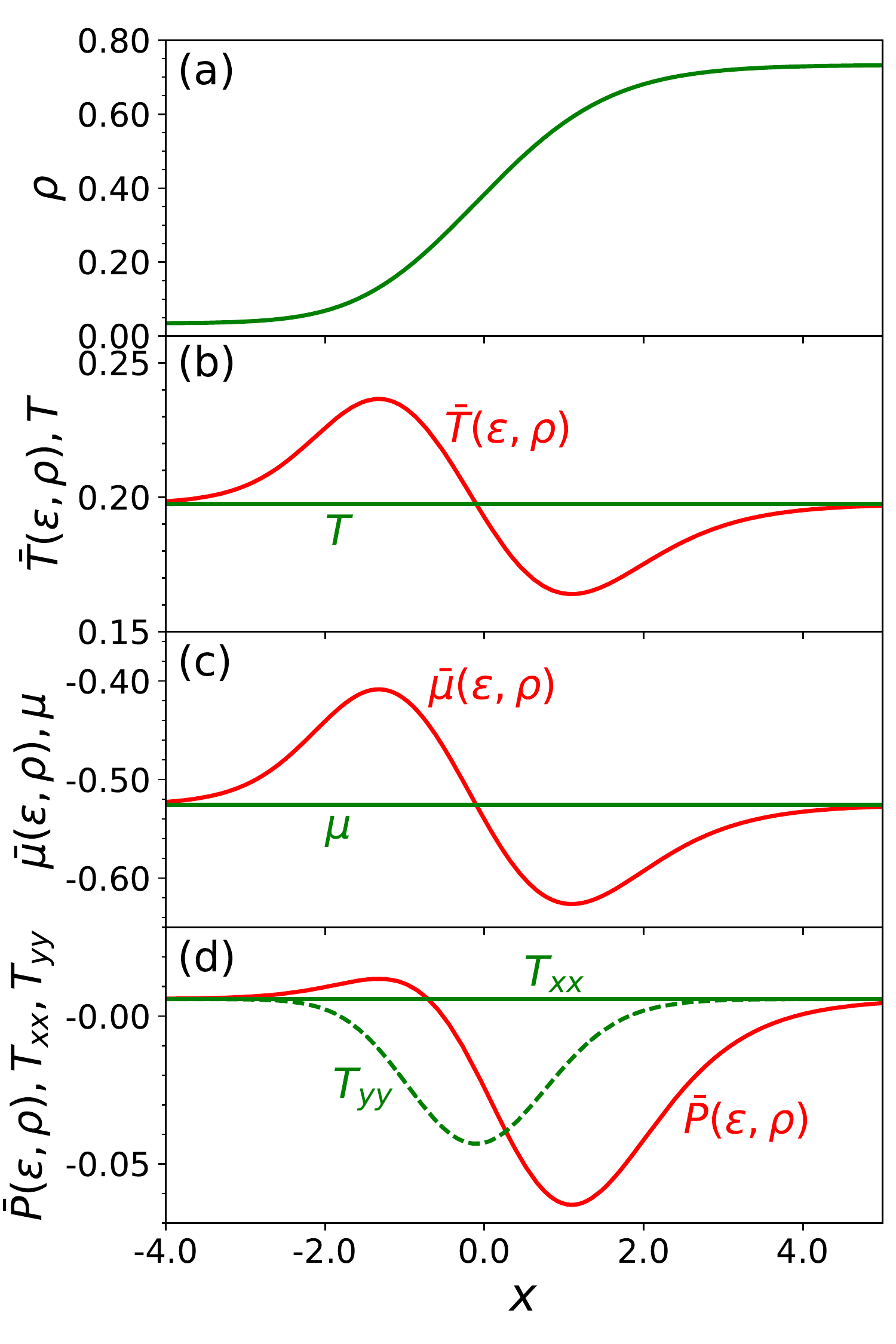}}
\caption{\label{fig:vdw}
In the upper panel (a), the density profile is displayed for the liquid-gas interface. Including gradient modifications, the temperature (panel b), chemical potential (panel c) and the stress-energy tensor element $T_{xx}$ (panel d) are all constant, as is required for an equilibrated system. The quantities evaluated as a function of the local energy density and local momentum density, $\bar{T}(\epsilon,\rho)$, $\bar{\mu}(\epsilon,\rho)$ and $\bar{P}(\epsilon,\rho)$ all vary substantially across the interface. The element $T_{yy}$ (panel a) also varies, but because it does not depend on $y$ does not affect the constraint $\partial_iT_{ij}=0$. To make all quantities dimensionless, densities are in units of $\rho_s$, the temperature and chemical potential are in units of $a\rho_s$ and the pressure is in units of $a\rho_s^2$. The coordinate $x$ is in units of $\sqrt{\kappa/a}$.}
\end{figure}

Figure \ref{fig:vdw} shows the density profile and local thermodynamic quantities for the case where the two phases are at a temperature $T=2T_c/3$, and for $a=1$, $\kappa=1$ and $\rho_s=1$. The temperature, $T=\bar{T}(\epsilon_\kappa,\rho)$ and chemical potential, $\mu=\bar{\mu}(\epsilon_\kappa,\rho)-\kappa\rho\nabla^2\rho/2$, indeed turn out to be constant as expected. With the surface normal being the $x$ direction, $T_{xx}=\bar{P}(\epsilon_\kappa,\rho)-\kappa\rho\nabla^2\rho+\kappa(\partial_x\rho)^2/2$ is also constant. Even though $T_{yy}$ varies as a function of $x$, it does not violate the constraint that $\partial_iT_{ij}=0$ for an equilibrated system. 

\section{Density-Density Correlations}
\label{sec:thermalflucs}

Another example of equilibrated density profiles is that for the density-density correlation function,
\begin{eqnarray}
C_{\rho\rho}(\bm{r})&\equiv&\langle\delta\rho(r=0)\delta\rho(\bm{r})\rangle,\\
\nonumber
\delta\rho(\bm{r})&\equiv&\rho(\bm{r})-\langle\rho(\bm{r})\rangle.
\end{eqnarray}
This integrates to the susecptibility. For a large volume $\Omega$,
\begin{eqnarray}
\int d^3r~C_{\rho\rho}(\bm{r})&=&\frac{1}{\Omega}\int d^3rd^3r'~C_{\rho\rho}(\bm{r}-\bm{r}')\\
\nonumber
&=&\frac{1}{\Omega}\langle(Q-\langle Q\rangle)^2\rangle=\chi.
\end{eqnarray}
An array of correlation functions can be considered,
\begin{eqnarray}
C_{\rho\alpha}(\bm{r})&=&\langle\delta\rho(r=0)\delta\alpha(\bm{r})\rangle,\\
\nonumber
C_{\rho\beta}(\bm{r})&=&\langle\delta\rho(r=0)\delta\beta(\bm{r})\rangle,\\
\nonumber
C_{\rho,Tij}(\bm{r})&=&\langle\delta\rho(r=0)\delta T_{ij}(\bm{r})\rangle,\\
\nonumber
C_{\rho\bar{\alpha}}(\bm{r})&=&\langle\delta\rho(r=0)\delta\bar{\alpha}(\epsilon(\bm{r}),\rho(\bm{r}))\rangle,\\
\nonumber
C_{\rho\bar{\beta}}(\bm{r})&=&\langle\delta\rho(r=0)\delta\bar{\beta}(\epsilon(\bm{r}),\rho(\bm{r}))\rangle,\\
\nonumber
C_{\rho\bar{P}}(\bm{r})&=&\langle\delta\rho(r=0)\delta\bar{P}(\epsilon(\bm{r}),\rho(\bm{r}))\rangle.
\end{eqnarray}
The latter three correlations involve the chemical potential, temperature and pressure as functions of the local energy density and charge density without gradient modifications. Unlike the implicit assumption used Sec. \ref{sec:theory} of the barred quantities being evaluated at $\epsilon_\kappa=\epsilon+\kappa\nabla^2\rho/2$, they are evaluated at $\epsilon$.

For stability as a function of time,
\begin{eqnarray}
\frac{d}{dt}C_{\rho\rho}(\bm{r},t)&=&\left\langle\left(\frac{d}{dt}\delta\rho(\bm{r}=0,t)\right)\delta\rho(\bm{r},t)\right\rangle
+\left\langle\delta\rho(\bm{r},t)\left(\frac{d}{dt}\delta\rho(\bm{r}=0,t)\right)\right\rangle.
\end{eqnarray}
Here, it is assumed the system is at constant temperature, and that the density changes only due to diffusion. From Eq. (\ref{eq:Fick}) and current conservation one finds a modified diffusion equation,
\begin{eqnarray}
\label{eq:diffusion}
\frac{d}{dt}\delta\rho(\bm{r},t)&=&D\nabla^2\delta\rho-\ell^2D(\nabla^2)^2\delta\rho(\bm{r},t),\\
\nonumber
\ell^2&=&\beta\kappa\chi,
\end{eqnarray}
and the correlation function obeys the relation,
\begin{eqnarray}
\frac{d}{dt}C_{\rho\rho}(\bm{r},t)&=&2D\nabla^2\left[\left(1-\ell^2\nabla^2\right)C_{\rho\rho}(\bm{r},t)\right].
\end{eqnarray}
For the correlation to be stable, $(d/dt)C_{\rho\rho}=0$, which requires
\begin{eqnarray}
\label{eq:ldef}
\left(1-\ell^2\nabla^2\right)C_{\rho\rho}(\bm{r},t)&=&0,~r\ne 0.
\end{eqnarray}
The last three relations only apply for $r\ne 0$ because Eq. (\ref{eq:diffusion}) should not be applied to $\delta\rho$ within the correlation unless the two positions, $\bm{r}$ and $\bm{r}'$, are separated sufficiently so that they evolve separately in the correlation function $\langle \delta\rho(\bm{r}')\delta\rho(\bm{r})\rangle$. Encapsulating the short-distance behavior with a delta function, the correlation function then has a form,
\begin{eqnarray}
C_{\rho\rho}(\bm{r})&=&(\chi-\chi_0)\frac{e^{-r/\ell}}{4\pi\ell^2r}+\chi_0\delta^3(\bm{r}).
\end{eqnarray}
This form satisfies the constraint that the correlation integrates to the susceptibility and describes any short-distance structure to the correlation with a parameter $\chi_0$. As $\kappa\rightarrow 0$, when gradient modifications disappear, both terms become delta functions and all the correlation is short-range. One example of short-range correlation is a gas, where the only correlation is between a particle and itself.

From Eq. (\ref{eq:summary}), at constant temperature the chemical potential obeys the relation,
\begin{eqnarray}
\delta\alpha&=&\delta\bar{\alpha}+\beta\kappa\nabla^2\delta\rho\\
\nonumber
&=&\frac{\partial\bar{\alpha}}{\partial\rho}\delta\rho+\beta\kappa\nabla^2\delta\rho\\
\nonumber
&=&-\frac{1}{\chi}\left(1-\ell^2\nabla^2\right)\delta\rho=0,
\end{eqnarray}
where the last relation used Eq. (\ref{eq:diffusion}). This implies that all the correlation functions with $\delta\alpha$ vanish for $r\ne 0$ because $\alpha$, like the temperature, does not fluctuate even though the density does fluctuate.  However, the chemical potential, $\bar{\alpha}(\epsilon,\rho)$, which ignores the gradient modifications, does fluctuate as does $\bar{\beta}(\epsilon,\rho)$ and $\bar{P}(\epsilon,\rho)$. Assuming small fluctuations (only one power in $\delta\rho$ from Eq. (\ref{eq:summary})), the various correlations listed previously can be determined from $C_{\rho\rho}$,
\begin{eqnarray}
C_{\rho\alpha}(\bm{r})&=&C_{\rho\beta}({\bm r})=0,\\
\nonumber
\delta T_{ij}&=&\frac{\partial\bar{P}}{\partial\bar{\alpha}}\frac{\partial\bar{\alpha}}{\partial\rho}\delta_{ij}
-\beta\kappa\rho\nabla^2\delta\rho\delta_{ij},\\
\nonumber
C_{\rho,Tij}&=&\delta_{ij}\frac{\rho}{\chi}\left[1-\beta\kappa\chi\nabla^2\right]C_{\rho\rho}(\bm{r})=0,\\
\nonumber
C_{\rho\bar{\alpha}}(\bm{r})&=&C_{\rho\rho}({\bm r})\frac{\partial\bar{\alpha}}{\rho}=\frac{1}{\chi}C_{\rho\rho}(\bm{r}),\\
\nonumber
C_{\rho\bar{\beta}}(\bm{r})&=&C_{\rho\rho}({\bm r})\frac{\partial\bar{\beta}}{\partial\rho}=\frac{1}{\chi_{QE}}C_{\rho\rho}(\bm{r}),\\
\nonumber
\chi_{QE}&=&\frac{1}{V}\left\langle(Q-\langle Q\rangle)(E-\langle E\rangle)\right\rangle,\\
\nonumber
\delta\bar{P}&=&\frac{\partial\bar{P}}{\partial\bar{\alpha}}\frac{\partial\bar{\alpha}}{\partial\rho}\delta\rho=\frac{\rho}{\beta\chi}\delta\rho,\\
\nonumber
C_{\rho\bar{P}}&=&\frac{\rho}{\beta\chi}C_{\rho\rho}(\bm{r}).
\end{eqnarray}
These expressions only consider the lowest order terms in $\delta\rho$. As was the case for the density profile, expressions involving gradient-corrected thermodynamic quantities vanish, while those calculated assuming uniform densities did not.

\section{Gradient Terms and Noisy Hydrodynamics}
\label{sec:noise}

The equations of hydrodynamics do not include any correlation functions. Nonetheless, correlations can be calculated within those equations by adding random noise terms to the equations \cite{Kapusta:2014dja,Young:2014pka,Ling:2013ksb,Pratt:2016lol,Gavin:2016hmv}. The noise generates event-by-event fluctuations, where each event involves an independent hydrodynamic calculation. One can add a noise term, $j^{\rm(n)}$, to the current in such a way that for a static system noise generates correlations consistent with the the charge susceptibility $\chi$. In the standard method, the generated correlation is short-range, at a scale set by the noise in the correlation. Here, we show how a local noise term, combined with the gradient modifications presented here, can lead to correlations with the length scale $\ell=\sqrt{\beta\kappa\chi}$, see Eq. (\ref{eq:diffusion}). This length scale diverges at the critical point in mean field theory. The modifications to the noise term derived here not only reproduces the desired mean field correlations, but does not require fine-tuning the strength of the noise at $T_c$. Rather, the divergences appear due to $\chi$ diverging, which is a function of the equation of state, rather than from adjusting the noise terms.

One can define a correlation function and solve for its time dependence,
\begin{eqnarray}
\label{eq:dCdt}
C(\bm{r})&=&\langle\rho(0)\rho(\bm{r})\rangle,\\
\nonumber
\partial_tC(\bm{r})&=&2D\nabla^2C(\bm{r})-2\sigma\kappa(\nabla^2)^2C(\bm{r})+[\partial_tC]^{\rm(n)},
\end{eqnarray}
where the last term comes from adding the $\bm{\nabla}\cdot j^{\rm(n)}$ term to the expression for $\partial_t\rho$. The $\bm{\nabla}$ terms refer to derivatives with respect to the relative coordinate $r$. Motivated by the Kubo relation for the conductivity \cite{Ling:2013ksb,Kapusta:2014dja,Pratt:2016lol},
\begin{eqnarray}
\label{eq:kubo}
\langle j_i^{\rm(n)}(0)j_k^{\rm(n)}(\bm{r})\rangle
&=&2\sigma T\delta_{ik}\delta^4(r) +2B\partial_i\partial_k\delta^4(r).
\end{eqnarray}
This last term does not appear in standard noise treatments. It integrates to zero, and is being multiplied by a currently unconstrained parameter $B$.

The time evolution of the noise term above then becomes
\begin{eqnarray}
\label{eq:dCdtn}
[\partial_tC]^{\rm(n)}(\bm{r})&=&
\frac{1}{\Delta t}\int_{-\Delta t}^0dt_1dt_2~\langle \bm{\nabla}_1\cdot\bm{j}^{\rm(n)}(\bm{r}_1) \bm{\nabla}_2\cdot\bm{j}^{\rm(n)}(\bm{r}_2)
\rangle\\
\nonumber
&=&-2\sigma T \nabla^2\delta^3(\bm{r}) +2B\nabla^2\nabla^2\delta^3(\bm{r}).
\end{eqnarray}

Next, we look for solutions for the steady-state case of the form,
\begin{equation}
\label{eq:Cform}
C(\bm{r})=\chi_0\delta^3(\bm{r})+(\chi-\chi_0)\frac{e^{-r/\ell}}{4\pi\ell^2r}.
\end{equation}
This ensures that the correlation integrates to the susceptibility, and $\chi_0$ is the strength of the short-range correlation. Using the fact that
\begin{equation}
\left(\nabla^2-\frac{1}{\ell^2}\right) \left(\frac{e^{-r/\ell}}{r}\right)=-4\pi\delta^3(\bm{r}),
\end{equation}
one can plug everything into Eq. (\ref{eq:dCdt}), using the facts that $\sigma=\beta D\chi$ and $\ell^2=\beta\kappa\chi$, to obtain
\begin{eqnarray}
\partial_tC(\bm{r})&=&0\\
\nonumber
&=&-2\left(\sigma\kappa\chi_0 -B\right) (\nabla^2)^2\delta^3(\bm{r}).
\end{eqnarray}
These determines the parameter $B$,
\begin{eqnarray}
B&=&\sigma\kappa\chi_0.
\end{eqnarray}
Thus, the noise as described in Eq. (\ref{eq:kubo}) is perfectly smooth near the phase transition if $\chi_0$ is smooth. The susceptibility $\chi$ diverges near $T_c$, which causes $\ell^2=\beta\kappa\chi$ to diverge and forces the diffusion constant, $D=\sigma T/\chi$, to vanish. The correlation function from Eq. (\ref{eq:Cform}) becomes
\begin{eqnarray}
C(\bm{r})&=&C(\bm{r})=\chi_0\delta^3(\bm{r})+\frac{(\chi-\chi_0)T}{\kappa\chi}\frac{e^{-r/\ell}}{4\pi}.
\end{eqnarray}
Thus, as one approaches the critical point the magnitude of the correlation approaches a constant, $T_c/(4\pi\kappa)$, and the correlation length diverges. The correlation length is also proportional to $\kappa$, which from the introduction is proportional to the range of the attractive interaction. If $\kappa$ is small, all the correlation is short-range. 

One can separate out the part of the correlation that is not short-range.
\begin{equation}
C'(\bm{r})\equiv C(\bm{r})-\chi_0\delta^3(\bm{r}),
\end{equation}
i.e. it is not the part of the correlation of a particle with itself, or some additional very-short range correlation. Plugging this into Eq.s (\ref{eq:dCdt}) and (\ref{eq:dCdtn}), 
\begin{eqnarray}
\label{eq:dCprimedt}
\partial_tC'(\bm{r},t)-2D\nabla^2C'(\bm{r},t)+2\sigma\beta\kappa(\nabla^2)^2C'(\bm{r},t)
&=&-2D(\chi-\chi_0)\nabla^2\delta^3(\bm{r})-\delta^3(\bm{r})\partial_t\chi_0(t)\\
\nonumber
\partial_tC'(\bm{r},t)-2D\nabla^2(1-\ell^2\nabla^2)C'(\bm{r},t)
&=&-2D(\chi-\chi_0)\nabla^2\delta^3(\bm{r})-\delta^3(\bm{r})\partial_t\chi_0(t).
\end{eqnarray}
Integrating the correlation,
\begin{eqnarray}
\frac{d}{dt}\int d^3r~C'(\bm{r},t)&=&-\frac{d}{dt}\chi_0.
\end{eqnarray}
This emphasizes that the integrated correlation is constant in time, and the rate of change of the short-range part of the susceptibility serves as a source term for a modified diffusion equation for $C'(\bm{r},t)$, similar as in \cite{Pratt:2016lol}, but with gradient terms modifying the diffusion equation. If the evolution is pursued indefinitely, the part of the correlation that cancels the equilibrium portion will spread over an infinitely large volume.

\section{Summary and Outlook}

Gradient terms have commonly appeared in dynamical theories of heavy-ion collisions, both in microscopic pictures where the mean field is altered, and even in hydrodynamic treatments where the pressure is altered. Here, a framework is presented providing consistent gradient terms for the stress-energy tensor, temperature, chemical potential, entropy density, and entropy current for a hydrodynamic treatment. This also determines how diffusion should be altered by gradient terms. The gradient terms in Eq.s (\ref{eq:summary}) are determined by one additional parameter $\kappa$, which accounts for the non-zero range attractive interaction between charged particles. These relations all resulted from the assumption of the form for the entropy density in Eq. (\ref{eq:Sdef}).

Here, a single kind of charge has been considered, whereas in a serious model of heavy-ion collisions one would expand the definition of charge density to include baryon, strangeness and isospin, or equivalently up, down and strange charge. It would not be difficult to expand the relations to include multiple charges. It would be more difficult to expand the expressions if the gradient terms were to include gradients of the energy density or if $\kappa$ were no longer a constant. For these more complicated considerations one would still begin with the assumption of a form for the entropy density as suggested in Eq. (\ref{eq:Sdefcomplicated}). For example, at zero baryon density one might expect a dipole-dipole attractive interaction in the quark-gluon plasma with a non-zero range between regions of higher energy density. This would motivate a term $\kappa_{\epsilon\epsilon}\epsilon\nabla^2\epsilon/2$. However, such terms are not likely to be as strong as the term involving only the charge, because those interactions involve charges directly rather than induced dipole moments. Surely, the parameter $\kappa_{\rho\rho}$ considered here could have some dependence of energy or charge density. If such terms were deemed important, the general approach applied here could be expanded.

The treatment of correlations from Sec.s \ref{sec:thermalflucs} and \ref{sec:noise} is based on the same assumption used to describe critical phenomena in Landau mean field theory. Hence, its validity near the critical point is questionable given that a three-dimensional system does not satisfy the Ginzburg criteria \cite{HuangStatMech}. If one is sufficiently close to the critical point, fluctuations of $\langle(\delta\rho)^3\rangle$ or $\langle(\delta\rho)^4\rangle$ overwhelm the corrections of $\langle(\delta\rho)^2\rangle$ considered here. However, this failure is only in the immediate region of the critical point, and even then does not qualitatively affect the behavior. More importantly, once inside the coexistence region, the seeding of fluctuations remains an open question, as the length scale of thermal noise, as in Sec. \ref{sec:noise}, may play a role. Currently, theory cannot rule out the possibility there is no phase transition, i.e. no coexistence region or critical point. If that is the case, but if susceptibility is large, significant non-zero-range correlations could develop and should be reasonably addressed with approaches similar to what is shown here.

The formalism here suffers from serious issues regarding causality. The stress-energy tensor has contributions proportional to $\nabla^2\delta\rho$, which clearly leads to the frequency of sound waves, $\omega$, obeying dispersion relations at high wave number $k$, with $\omega^2\sim k^4$, which at some point becomes super-luminal. This would be important for systems with small-scale density inhomogeneities characterized by high wave numbers. Israel-Stewart hydrodynamics successfully addresses similar issues that occur when adding viscosity to ideal hydrodynamics \cite{Muronga:2003ta,Muronga:2001zk}. Similar ideas might prove successful for non-zero-range corrections. Another place where causality plays a role is with the treatment of diffusion, Eq. (\ref{eq:diffusion}). Even without gradient modifications, the diffusion equation is parabolic, $\omega\sim k^2$, and gives acausal behaviors at short times where the spread from a point source extends a distance, $\sqrt{2Dt}$, which is greater than $ct$ for short times. A number of approaches have addressed this issue \cite{Gavin:2016jfw,Gavin:2016nir,Kapusta:2014dja,Cattaneo,GurtinPipkin,GurtinPipkinB}. In Eq. (\ref{eq:diffusion}), $\omega\sim k^4$ for large $k$, which suggests even  more unphysical behavior for short times.  Similar tactics should be developed to limit the diffusive spread at small times.

Despite the issues listed above, the formalism developed here should prove useful. In a dynamic environment, such as a heavy-ion collision, density correlations develop and grow both hydrodynamically and diffusively, and both forms of growth are necessary for a system to equilibrate. This formalism allows the simultaneous and consistent treatment of both. In future work, the author plans to apply these techniques to study the growth of fluctuations, and discern whether the features of the phase diagram near the critical point might manifest themselves in final-state measurements. Given the rapid nature of the expansion of the fireball in heavy-ion collisions, it is not clear that sufficient time exists for the development of correlations that can be detected experimentally. Thus, it is critical to develop dynamic models for the growth of fluctuations.

\begin{acknowledgments}
This work was supported by the Department of Energy Office of Science through grant number DE-FG02-03ER41259. The author thanks Chris Plumberg, Joe Kapusta, Volker Koch, Marcus Bluhm and Yi Yin for stimulating discussions.
\end{acknowledgments}


\end{document}